\documentclass[useAMS,usenatbib]{mn2e}
\usepackage{graphicx}
\usepackage{natbib}
\usepackage{amsmath}
\usepackage{color}
\usepackage{arydshln}
\usepackage{tabularx}
\usepackage{colortbl}
\usepackage{hhline}

\usepackage{mathrsfs}

\usepackage{algorithm}
\usepackage[noend]{algpseudocode}

\makeatletter
\def\BState{\State\hskip-\ALG@thistlm}
\makeatother

\usepackage{array}
\usepackage{multirow}

\newcommand\MyBox[2]{
  \fbox{\lower0.75cm
    \vbox to 1.0cm{\vfil
      \hbox to 2.0cm{\hfil\parbox{1.4cm}{#1\\#2}\hfil}
      \vfil}%
  }%
}

\usepackage{placeins} 

\newcommand{\url}{{url\;}}

\title[Smooth stochastic density field reconstruction]{Smooth stochastic density field reconstruction}
\author[Aragon-Calvo M.A.]{M.A. Aragon-Calvo$^{1}$ \thanks{E-mail:maragon@astro.unam.mx} \\
$^{1}$Instituto de Astronom\'{i}a, UNAM, Apdo. Postal 106, Ensenada 22800, B.C., M\'{e}xico\\}
\begin{document}

\date{}

\pagerange{\pageref{firstpage}--\pageref{lastpage}} \pubyear{2002}
\maketitle
\label{firstpage}

\begin{abstract}

We introduce a method for generating a continuous, mass-conserving and high-order differentiable density field from a discrete point distribution such as particles or halos from an N-body simulation or galaxies from a spectroscopic survey. The method consists on generating an ensemble of point realizations by perturbing the original point set following the geometric constraints imposed by the Delaunay tessellation in the vicinity of each point in the set.  By computing the mean field of the ensemble we are able to significantly reduce artifacts arising from the Delaunay tessellation in poorly sampled regions while conserving the features in the point distribution. Our implementation is based on the Delaunay Tessellation Field Estimation  (DTFE) method, however other tessellation techniques are possible. 
The method presented here shares the same advantages of the DTFE method  such as self-adaptive scale, mass conservation and continuity, while being able to reconstruct even the faintest structures of the point distribution usually dominated by artifacts in Delaunay-based methods. 

Additionally, we also present preliminary results of an application of this method to image denoising and artifact removal, highlighting the broad applicability of the technique introduced here.

\end{abstract}

\section{Introduction}\label{sec:intro}

Computing a continuous density field from a discrete point set is a common task for problems that require the use of grid-based methods. This is a non-trivial task and no existing approach offers a perfect solution. The choice of a particular density estimation method is often a compromise based on the nature of the data at hand, between the desired properties of the density field and the limitations of the method. The simplest density estimation approach is to place points inside a grid, effectively computing an n-dimensional histogram. This approach, while straightforward to implement and interpret, requires a good sampling per cell. Other more sophisticated kernel-based methods include, Cloud in Cell (CIC), Triangle In Cell and Gaussian kernel (see \citet{Aarseth08,Martinez09} for an introduction on density estimation methods in cosmology). In general these methods are dominated by shot noise in poorly sampled regions strongly limiting their application to sparsely sampled datasets such as the galaxy distribution (see  also \citet{Schaap00} for an excellent review).

A major improvement in the reconstruction of density fields from point sets was introduced by \citet{Schaap00, Schaap07} with the Delaunay Tessellation Field Estimator method (DTFE). The DTFE method assumes a relation between the volume of the adjacent Voronoi cell of a given point and its density. From the density estimates the Delaunay tessellation itself can be used to produce a continuous linearly interpolated density field. This is usually done on a regular grid although other sampling schemes are possible. The DTFE method inherits the self-adaptive scaling, volume filling and anisotropic nature of the Delaunay tessellation allowing it to closely follow the intricate patterns in the distribution of matter. The linear interpolation in the DTFE method can be replaced by higher order schemes such as in the natural neighbours approach \citep{Sibson81, Watson92}. While natural neighbours produce smooth density fields it does not conserve mass and therefore its use is limited to applications based only on the geometry of the features in the density field (see \citet{Platen11} for a more complete discussion). Other Tessellation-based density estimation and interpolation methods exist that share some of the properties of the Delaunay-based methods such as the Lagrangian sheet approach of \citep{Abel12,Shandarin12}  limited to particles in N-body simulations which produces unparalleled results in low and medium density regions but severe artifacts in dense regions.

A well-known  problem of Delaunay-based methods is their tendency to produce high-density spikes in cases where a given point is connected to a dense clusters of points  separated by a distance significantly larger than the mean inter-particle separation. Delaunay-based methods are also sensitive to small perturbations in the point distribution which can result in large rearrangements in the tessellation, thus potentially limiting their application in time-evolving point distributions. These artifacts are particularly problematic for methods that study the morphology and topology of the structures in the observed galaxy distribution where we do not know the exact position of the galaxies and small position errors can results in significantly different density estimations. In order to alleviate some of the artifacts introduced by the Delaunay tesselation a common approach is to apply a smoothing procedure \citep{Aragon07,Aragon10, Platen11}, partly removing the advantage of self-adaptive scale and mass conservation.

\begin{figure}
  \centering
  \includegraphics[width=0.4\textwidth,angle=0.0]{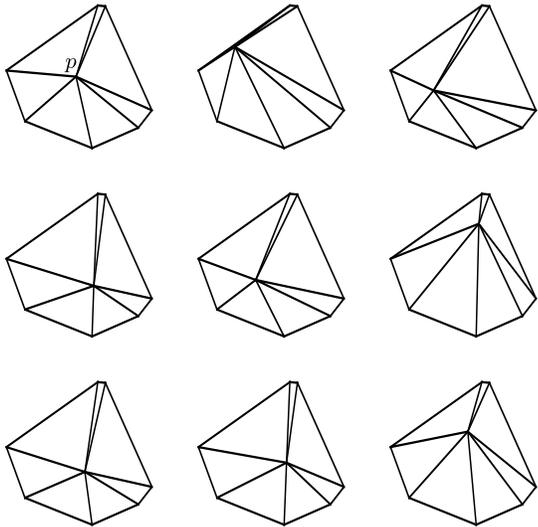}
  \caption{Adjacent Voronoi cell of point $p$ (top left sub-figure) and 8 random perturbations (remaining sub-figures) with their corresponding Delaunay Tessellation. Note that only the target point is perturbed.}\label{fig:random_perturbations}
\end{figure}

%
\section{Self-adaptive Stochastic perturbations}

The method presented here is based on the observation that, in the case of coherent structures, the density field inside a Voronoi Cell follows the points describing the cell itself \citep{Aragon07}. This property can be exploited to generate new sampling points inside the cell that naturally follow the anisotropies in the point distribution. For a given point set $P$ from which we computed its Delaunay tessellation we generate an ensemble of perturbed point sets as:

\begin{equation}\label{eq:perturbation} 
P^i = P +N^i
\end{equation}

\noindent where $P^i$ is the perturbed $i^{\textrm{\tiny{th}}}$ realization in the ensemble and $N^i$ is a random perturbation. In order to conserve the scale and anisotropy of the features present in $P^i$ we constraint the perturbed point $p_j$ to lie inside the adjacent Voronoi cell $\mathscr{V}_j$ of the original point $p$ (see Fig. \ref{fig:random_perturbations}). Without this constraint the perturbation would act as an isotropic filter, removing some of the advantages of the Delaunay tessellation.

\begin{figure}
  \centering
  \includegraphics[width=0.5\textwidth,angle=0.0]{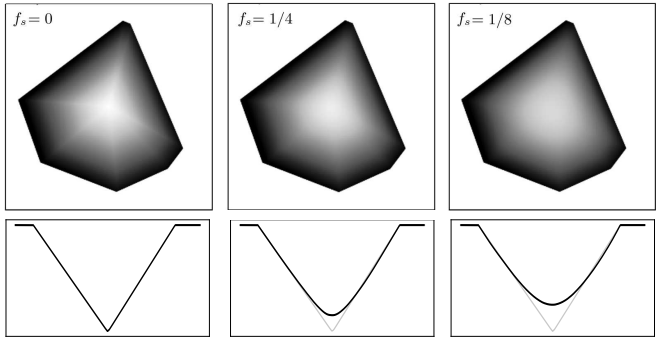}
  \caption{Effect of the dispersion scaling factor $f_s$ in the random sampling scheme applied to a point at the center of its adjacent Voronoi cell. The top panels show the interpolated DTFE density field inside  $\mathscr{V}$ and the bottom panels show a linear profile across the center of  $\mathscr{V}$ along the $x$-axis. From left to right we show the SDTFE fields computed by  perturbing the central point by with Gaussian noise with dispersion equal to $f_s=0$ (unperturbed point), $1/8$ and $1/4$ of the equivalent radius of $\mathscr{V}$ respectively. Note that in this example the points defining $\mathscr{V}$ remain fixed at their original position and all have the same density.}\label{fig:voronoi_samplings}
\end{figure}

%
\subsection{Smoothing kernel function}

The probability distribution from which the random perturbation $N$ is drawn must ideally reflect the nature of the underlying density field being sampled. In principle we do not know the true density field sampled by the point set. However, given the self-adaptive nature of the Delaunay/Voronoi tessellations we can assume that the tessellation itself gives us a good indication of the scale and shape of the structures in the density field. In the simplest case $N$ can be drawn from a uniform Poisson sampling. This option may be acceptable when we do not have more information available on the density field than the sample $P$ itself. Biased samples such as galaxies populate regions of enhanced formation likelihood and could be better described as a thresholded sample or a localized function such as a Gaussian distribution. In this work we draw random points from a Gaussian distribution: $p_j  \in G(\sigma_j)$ where the dispersion $\sigma_j$ is equal to the equivalent radius of $\mathscr{V}_j$:

\begin{equation}
\sigma_j = \left ( \frac{3}{4 \pi} |  \mathscr{V}_j | \right )^{1/3}.
\end{equation}

\noindent We introduce a scaling factor $f_s$ that controls the scale $\sigma_j$ as:

\begin{equation}
\hat{\sigma_j} = f_s \; \sigma_j .
\end{equation}

\noindent Values of $f_s$ close to unity produce smooth fields while smaller values produce fields tending to the limiting case of the original point distribution. The factor $f_s$ can be considered a pseudo free-parameter since it has a small effect in the  the quality of the density reconstruction for values close to 1 and can be safely omitted. We kept it in order to fine-tune the reconstruction.  If a perturbation brings the new sampling point outside $\mathscr{V}_j$ we repeat the process in equation \ref{eq:perturbation} until the point lies inside $\mathscr{V}_j$.  Note that while the perturbations are isotropic the volume constraint is not, conforming the sampling function to the shape of the adjacent Voronoi cells. The procedure above is described in table \ref{alg:pseudo-code}.

For each realization $P^i$ we compute its continuous density field by interpolating on a regular grid following the standard DTFE method and then proceed to compute the mean of the whole ensemble. By computing the mean of the ensemble we ``average out" artifacts arising from the Delaunay tessellation in individual realizations. The mean density field shares the same properties as the single-realization DTFE such as anisotropy, self-scaling and mass conservation.  Given that our implementation of stochastic sampling is based on the DTFE interpolation scheme we refer to it as  Stochastic Delaunay Tessellation Field estimator (SDTFE). However, note that the method is general other tessellation and interpolation schemes are possible.

Figure \ref{fig:voronoi_samplings} shows the stochastic sampling of the interior of the adjacent Voronoi cell $\mathscr{V}_j$ of point $p$. Compare the linear artifacts and discontinuities present in the DTFE field with the smooth field computed with SDTFE. Note that in this example the density estimates at the position of the points in the periphery of $\mathscr{V}_j$ is constant. In a real tessellation the density will be different and will reflect the local anisotropies which will further increase the artifacts inside $\mathscr{V}_j$. While Fig.  \ref{fig:voronoi_samplings} seems to indicate lack of mass conservation (since the integral of the two SDTFE examples seems higher that the DTFE case) this is compensated in other regions of $\mathscr{V}_j$ as numerically confirmed from a large ensemble.

%
\begin{algorithm}
\caption{Stochastic DTFE}\label{alg:pseudo-code}
\begin{algorithmic}[1]

\State Read point set $P$
\State Compute Delaunay tessellation from $P$
\State Compute $\sigma_i$ and $\hat{\sigma_i}$ for each point
    
\For {each realization $i$}
	\For {each point $j$}
		\While {$p_j$ outside $\mathscr{V}_j$}
		\State Generate random perturbation $N^i_j$		
		\EndWhile
	\EndFor
	
	\State Compute DTFE field from $P^i$ and store
\EndFor

\State Compute mean from ensemble
\end{algorithmic}
\end{algorithm}

\begin{figure*}
  \centering
  \includegraphics[width=0.99\textwidth,angle=0.0]{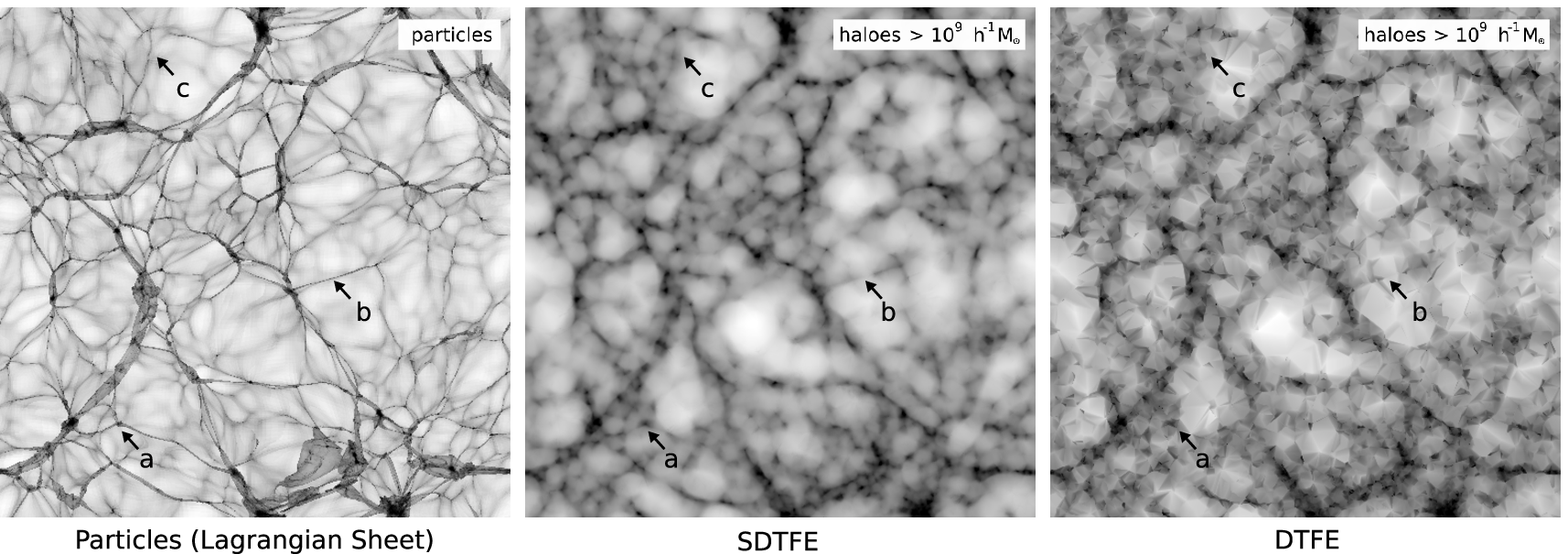}
  \caption{Density reconstruction from a sample of subhaloes more massive than $10^9 h^{-1} M_{\odot}$ in the Illustris simulation. Left panel shows (as a reference) the density field computed from the particle distribution using the Lagrangian Sheet density estimation method \citep{Abel12,Shandarin12}. Center and right panels show the SDTFE and  DTFE density field reconstructions respectively. Three illustrative regions are labeled (see text for details).}\label{fig:illustris}
\end{figure*}

%
\section{Data}\label{sec:data}

The results presented in this work are based on an N-body dark matter computer simulation and a galaxy redshift survey. The dark matter simulation was obtained from the Illustris project\footnote{www.illustris-project.org} which consists of a set of N-body hydrodynamical and dark- matter-only computer simulations ran on a 106.5 Mpc box with the concordance cosmological parameters: $\Omega_m = 0.2726, \Omega_{\Lambda}=0.7274, h=0.704$ \citep{Vogelsberger14}. We ran a custom simulation resampled from the reconstructed initial conditions into a $300^3$ particle grid. The subhalo catalog used for our analysis was extracted from the dark matter-only Dark-3 run available at the Illustris project's website.

In addition to the N-body simulation we used a galaxy catalog extracted from the Sloan Digital Sky Survey\footnote{www.sdss.org} (SDSS, \citet{York00,Abazajian09}). We selected a sample of galaxies with redshifts measured spectroscopically in the range $z=0.01-0.12$. We compressed redshift distortions in dense clusters and groups (Fingers of God) by identifying groups of galaxies elongated along the line of sight and compressing them to match their dispersion in the plane of the sky and along the line of sight. The details of the dataset and post-processing are discussed in \citet{Aragon15}.

%
\section{Results}

In this section we present the results of our method applied to a sample of subhaloes extracted from the Illustris simulation \citep{Vogelsberger14}  and the catalog of galaxies extracted from the SDSS project \citep{York00} described in Sec. \ref{sec:data}.

Figure \ref{fig:illustris} shows a comparison between the subhalo density field computed using SDTFE and the original DTFE method. The point sample corresponds to the center of mass of subhaloes with $M > 10^9 h^{-1} M_{\odot}$. There are $660607$ haloes in that mass range in the simulation box giving a density of haloes of 1.56 $h^{3}$Mpc$^{-3}$, roughly of the same order as the observed galaxy distribution. For simplicity we did not include the mass of the haloes in the density estimation. As a reference we also show in Fig. \ref{fig:illustris} the density field computed from the dark matter particles in a low resolution run with  $300^3$ particles using the  Lagrange Tessellation technique described in \citep{Abel12,Shandarin12}. The density of particles is $\sim 64 h^{-1}$Mpc$^{-3}$. 
Visual inspection shows that the density field computed with SDTFE contains less artifacts than the field computed with the original DTFE while retaining the original features in the density field. This is particularly clear in  low-medium density regions where the Delaunay tessellation is dominated by prominent highly elongated spike-like tetrahedra. The density field inside high density regions is practically the same for both SDTFE and DTFE. For comparison we highlight three regions in the density slice shown:

Region (a) corresponds to a halo sitting at the intersection of three filaments near the edge of a void. The filaments are seen in the particle distribution as thin bridges connecting to the halo. The SDTFE field is able to reconstruct the two bottom filaments and the top filament (although is seems to split in two structures) while the DTFE field in the vicinity of the halo is dominated by Delaunay artifacts and barely contains any coherent structure. 

Region (b) corresponds to a tenuous intra-void wall seen edge-on. This is a challenging task given the low density of haloes in void environments and this is reflected in the gap of haloes near the center of the wall seen in both DTFE and SDTFE fields. This is an example of a low-density region dominated by high-density Delaunay spikes-like artifacts. The wall in the DTFE field has several perpendicular spikes pointing to the interior of its adjacent voids. This is not seen in the particle distribution and is purely an effect of the Delaunay tessellation connecting the haloes in the wall with other haloes inside or at the border of the void. In contrast the spikes are much less prominent (although still present) in the SDTFE field. 

Region (c) corresponds to a complex system of tenuous walls and filaments inside a void. Here the artifacts introduced by the Delaunay tessellation are more significant given the low density and complexity of the region. The DTFE field contains no visible coherent structure, being completely dominated by Delaunay artifacts. This is in contrast to the SDTFE field which closely delineates the tenuous structures in the region. It is remarkable that the SDTFE can reconstruct the density field with such a low density of sampling points. Paradoxically the perturbing scheme in the  SDTFE method takes advantage of the same ability of the Delaunay tessellation to follow local anisotropies that introduce artifacts in the DTFE field.


\begin{figure*}
  \centering
  \includegraphics[width=0.95\textwidth,angle=0.0]{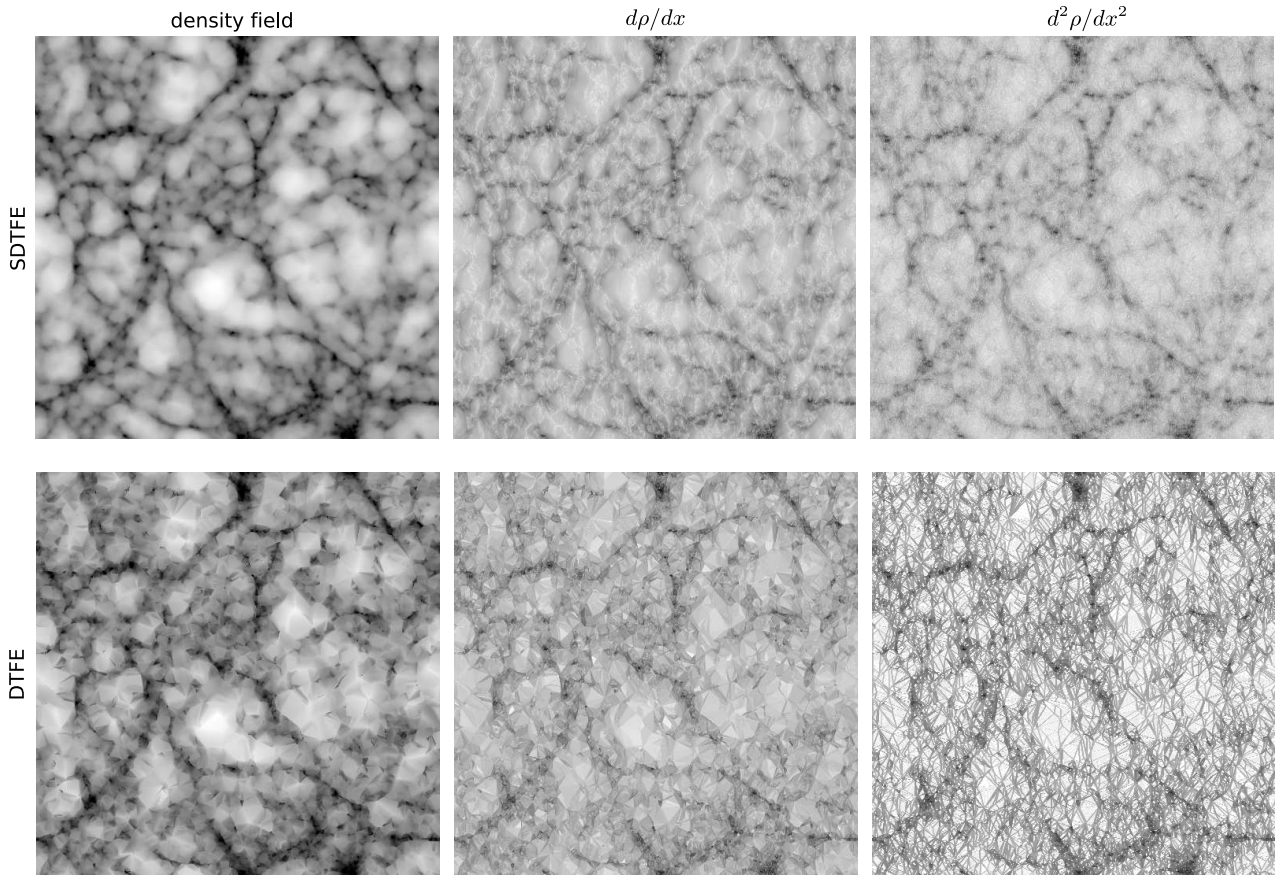}
  \caption{First and second-order derivatives of the density field. Top panels were computed from the SDTFE field while lower panels show the results computed from the  DTFE field. The derivatives correspond to the density field slice shown in Fig. \ref{fig:illustris}. Left panels show the first derivative $d\rho/dx$ and right panels show the second derivative $d^2 \rho / d x^2$ of the density field. Note that since DTFE uses linear interpolation its second derivative is zero except at the boundaries between tetrahedra. }\label{fig:illustris_derivs}
\end{figure*}

\subsection{First and second-order derivatives}

Figure \ref{fig:illustris_derivs} shows a comparison between the first and second derivatives of the density field computed with the DTFE and SDTFE methods. The first-order derivatives of the DTFE field are constant inside individual tetrahedra as a result of the linear interpolation scheme and are discontinuous at the boundaries between tetrahedra. In contrast the first derivatives of SDTFE smoothly change across the full tessellation. The contrast between the methods is larger for the second-order derivatives where the SDTFE field is still smooth and can be further differentiated while the DTFE field is discontinuous. 

\begin{figure*}
  \centering
  \includegraphics[width=0.8\textwidth,angle=0.0]{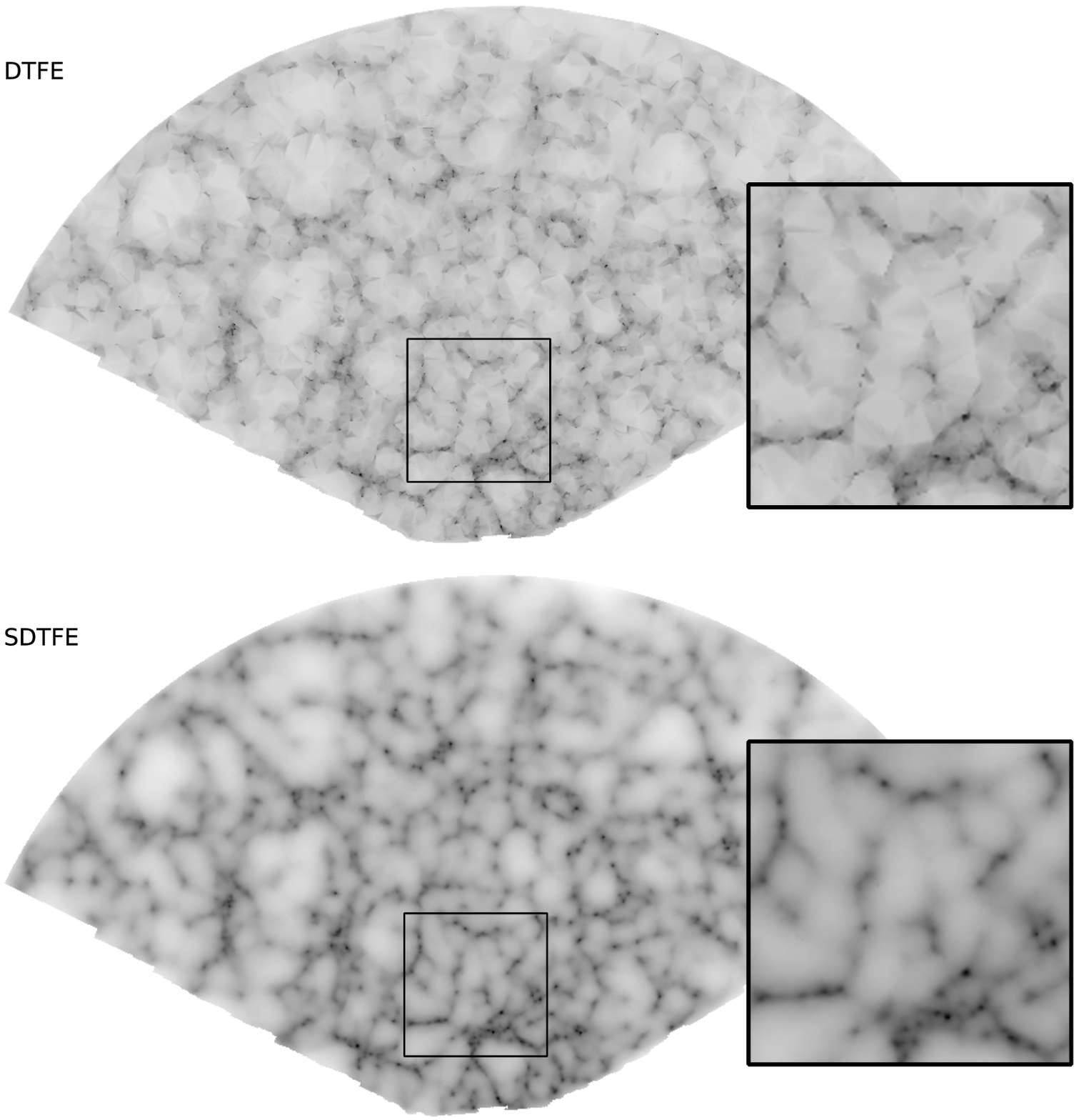}
  \caption{Comparison between the density field computed with the standard DTFE (top) and the SDTFE method (bottom) from the galaxy distribution observed by the SDSS survey. The image shows a low declination slice across the survey in the redshift range $z=0.01-0.12$. The subplots show a zoom-in into a complex LSS region containing several galaxy groups, filaments, walls and voids. The density field was rescaled for clarity.}\label{fig:sdss}
\end{figure*}

%
\subsection{SDSS}

Figure \ref{fig:sdss} shows a comparison between the galaxy density field computed with the DTFE and SDTFE methods. The galaxy distribution observed by the SDSS survey is a strongly biased sample of the underlying matter distribution (as in the case of the halo sample shown in Fig. \ref{fig:illustris}) resulting in Delaunay artifacts in the low-medium density regions. Voids in the DTFE field are completely dominated by Delaunay artifacts. The individual tetrahedra connected to the borders of the voids can be seen as well as denser spikes. On the other hand the SDTFE field is smooth and has few artifacts. 

The  zoom-in region in fig. \ref{fig:sdss}  further highlights the differences between the two methods in the reconstruction of poorly sampled  complex geometries. The SDTFE closely follows the anisotropic features in the galaxy distribution while smoothly interpolating the density field between galaxies. Note the degradation in the quality of the DTFE density reconstruction with increasing redshift as the density of galaxies decreases due to the galaxy selection function: distant regions of the survey ($z \sim 0.1)$ show few coherent structures compared to nearby regions. In comparison, there is less dependence in the quality of the identified structures with distance in the density field computed with the SDTFE method. Even at the most distant regions the density field shows the full anisotropic nature of the cosmic web. 

%
\section{Discussion an Future Work}

We presented a new method to produce smooth and high-order differentiable density fields from a discrete point distribution. The implementation presented here is based on the Delaunay Tessellation Field Estimator method and shares its advantages, namely self-adaptive scale, mass conservation, continuity and high-order differentiability. Our method can be considered a smoothing scheme in which the smoothing kernel is defined by the point distribution from which the density field is reconstructed. Here the adjacent Voronoi tessellation of the point distribution defines the scale and shape of a filter to be applied locally. As such it is conceptually similar to other self-adaptive smoothing techniques such as anisotropic Gaussian filtering without being limited by grid resolution. The proposed method is able to reconstruct even the faintest structures in the point distribution usually dominated by artifacts in standard Delaunay-based methods. The high-order differenciable property of SDTFE makes it an ideal candidate for gradient-based LSS analysis methods \citep{Aragon07, Platen07,Aragon10b}.

The method presented here is easy to implement on top of existing DTFE pipelines as it only requires changing the point distribution. Being and ensemble method it is straightforward to parallelize. Computation time can potentially be significantly decreased  by reusing the Delaunay tessellation of the original point set as a basis to compute the rest of the ensemble since, by construction,  every perturbed point lies inside its original adjacent Voronoi cell.

It is interesting to note that, as our method shows,  the Delaunay tessellation contains the information needed to produce an accurate reconstruction of the structures sampled by the point distribution. The unwanted artifacts in the original DTFE method are partly the result of the linear interpolation scheme. Other non-linear interpolations such as natural-neighbours produce smooth reconstructions similar to the one presented here but do not conserve mass while being significantly more complex to implement and computationally expensive.

The idea of generating an ensemble from a constrained perturbed point distribution can be applied to other problems. In appendix \ref{app:image} we present preliminary results of this idea applied to image denoising and artifact removing. This approach, while computationally intensive compared to other image processing algorithms, is straightforward to interpret and produces excellent results. We can also use ensembles to improve methods for characterizing of the Cosmic Web. In a future paper we will present an ensemble-based implementation of the Spine method \citep{Aragon10b} which is robust against noise and can give a direct indication of the significance of the identified structures.

%
\section{Acknowledgments}
The author would like to thank Rien van de Weygaert for discussions and comments on the method. This work was partly funded by ``Programa de Apoyo a Proyectos de Investigaci\'{o}n e Innovaci\'{o}n Tecnol\'{o}gica'' grant DGAPA-PAPIIT IA104818.

\bibliography{refs} 

\begin{thebibliography}{17}
\expandafter\ifx\csname natexlab\endcsname\relax\def\natexlab#1{#1}\fi

\bibitem[{{Aarseth}, {Tout} \& {Mardling}(2008){Aarseth}, {Tout}, \&
  {Mardling}}]{Aarseth08}
{Aarseth} S.~J., {Tout} C.~A., {Mardling} R.~A., 2008, {The Cambridge N-Body
  Lectures}, Vol. 760

\bibitem[{{Abazajian} {et~al}\mbox{.}(2009){Abazajian}, {Adelman-McCarthy},
  {Ag{\"u}eros}, {Allam}, {Allende Prieto}, {An}, {Anderson}, {Anderson},
  {Annis}, {Bahcall}, \& et~al.}]{Abazajian09}
{Abazajian} K.~N. {et~al.}, 2009, \apjs, 182, 543

\bibitem[{{Abel}, {Hahn} \& {Kaehler}(2012){Abel}, {Hahn}, \&
  {Kaehler}}]{Abel12}
{Abel} T., {Hahn} O., {Kaehler} R., 2012, \mnras, 427, 61

\bibitem[{{Arag{\'o}n-Calvo} {et~al}\mbox{.}(2007){Arag{\'o}n-Calvo}, {Jones},
  {van de Weygaert}, \& {van der Hulst}}]{Aragon07}
{Arag{\'o}n-Calvo} M.~A., {Jones} B.~J.~T., {van de Weygaert} R., {van der
  Hulst} J.~M., 2007, \aap, 474, 315

\bibitem[{{Arag{\'o}n-Calvo} {et~al}\mbox{.}(2010){Arag{\'o}n-Calvo}, {Platen},
  {van de Weygaert}, \& {Szalay}}]{Aragon10b}
{Arag{\'o}n-Calvo} M.~A., {Platen} E., {van de Weygaert} R., {Szalay} A.~S.,
  2010, \apj, 723, 364

\bibitem[{{Aragon-Calvo} {et~al}\mbox{.}(2010){Aragon-Calvo}, {van de
  Weygaert}, {Araya-Melo}, {Platen}, \& {Szalay}}]{Aragon10}
{Aragon-Calvo} M.~A., {van de Weygaert} R., {Araya-Melo} P.~A., {Platen} E.,
  {Szalay} A.~S., 2010, \mnras, 404, L89

\bibitem[{{Aragon-Calvo} {et~al}\mbox{.}(2015){Aragon-Calvo}, {van de
  Weygaert}, {Jones}, \& {Mobasher}}]{Aragon15}
{Aragon-Calvo} M.~A., {van de Weygaert} R., {Jones} B. J.~T., {Mobasher} B.,
  2015, \mnras, 454, 463

\bibitem[{{Mart{\'\i}nez} {et~al}\mbox{.}(2009){Mart{\'\i}nez}, {Saar},
  {Mart{\'\i}nez-Gonz{\'a}lez}, \& {Pons-Border{\'\i}a}}]{Martinez09}
{Mart{\'\i}nez} V.~J., {Saar} E., {Mart{\'\i}nez-Gonz{\'a}lez} E.,
  {Pons-Border{\'\i}a} M.~J., 2009, {Data Analysis in Cosmology}, Vol. 665

\bibitem[{{Platen}, {van de Weygaert} \& {Jones}(2007){Platen}, {van de
  Weygaert}, \& {Jones}}]{Platen07}
{Platen} E., {van de Weygaert} R., {Jones} B. J.~T., 2007, \mnras, 380, 551

\bibitem[{{Platen} {et~al}\mbox{.}(2011){Platen}, {van de Weygaert}, {Jones},
  {Vegter}, \& {Calvo}}]{Platen11}
{Platen} E., {van de Weygaert} R., {Jones} B. J.~T., {Vegter} G., {Calvo} M.
  A.~A., 2011, \mnras, 416, 2494

\bibitem[{{Schaap}(2007)}]{Schaap07}
{Schaap} W.~E., 2007, PhD thesis, Kapteyn Astronomical Institute

\bibitem[{{Schaap} \& {van de Weygaert}(2000)}]{Schaap00}
{Schaap} W.~E., {van de Weygaert} R., 2000, \aap, 363, L29

\bibitem[{{Shandarin}, {Habib} \& {Heitmann}(2012){Shandarin}, {Habib}, \&
  {Heitmann}}]{Shandarin12}
{Shandarin} S., {Habib} S., {Heitmann} K., 2012, \prd, 85, 083005

\bibitem[{{Sibson}(1981)}]{Sibson81}
{Sibson} R., 1981, Barnet V.(ed.), Wiley, Chichester

\bibitem[{{Vogelsberger} {et~al}\mbox{.}(2014){Vogelsberger}, {Genel},
  {Springel}, {Torrey}, {Sijacki}, {Xu}, {Snyder}, {Bird}, {Nelson}, \&
  {Hernquist}}]{Vogelsberger14}
{Vogelsberger} M. {et~al.}, 2014, \nat, 509, 177

\bibitem[{{Watson}(1992)}]{Watson92}
{Watson} D.~F., 1992, Pergamon Press

\bibitem[{{York} {et~al}\mbox{.}(2000){York}, {Adelman}, {Anderson},
  {Anderson}, {Annis}, {Bahcall}, {Bakken}, {Barkhouser}, {Bastian}, {Berman},
  {Boroski}, {Bracker}, {Briegel}, {Briggs}, {Brinkmann}, {Brunner}, {Burles},
  {Carey}, {Carr}, {Castander}, {Chen}, {Colestock}, {Connolly}, {Crocker},
  {Csabai}, {Czarapata}, {Davis}, {Doi}, {Dombeck}, {Eisenstein}, {Ellman},
  {Elms}, {Evans}, {Fan}, {Federwitz}, {Fiscelli}, {Friedman}, {Frieman},
  {Fukugita}, {Gillespie}, {Gunn}, {Gurbani}, {de Haas}, {Haldeman}, {Harris},
  {Hayes}, {Heckman}, {Hennessy}, {Hindsley}, {Holm}, {Holmgren}, {Huang},
  {Hull}, {Husby}, {Ichikawa}, {Ichikawa}, {Ivezi{\'c}}, {Kent}, {Kim},
  {Kinney}, {Klaene}, {Kleinman}, {Kleinman}, {Knapp}, {Korienek}, {Kron},
  {Kunszt}, {Lamb}, {Lee}, {Leger}, {Limmongkol}, {Lindenmeyer}, {Long},
  {Loomis}, {Loveday}, {Lucinio}, {Lupton}, {MacKinnon}, {Mannery}, {Mantsch},
  {Margon}, {McGehee}, {McKay}, {Meiksin}, {Merelli}, {Monet}, {Munn},
  {Narayanan}, {Nash}, {Neilsen}, {Neswold}, {Newberg}, {Nichol}, {Nicinski},
  {Nonino}, {Okada}, {Okamura}, {Ostriker}, {Owen}, {Pauls}, {Peoples},
  {Peterson}, {Petravick}, {Pier}, {Pope}, {Pordes}, {Prosapio},
  {Rechenmacher}, {Quinn}, {Richards}, {Richmond}, {Rivetta}, {Rockosi},
  {Ruthmansdorfer}, {Sandford}, {Schlegel}, {Schneider}, {Sekiguchi}, {Sergey},
  {Shimasaku}, {Siegmund}, {Smee}, {Smith}, {Snedden}, {Stone}, {Stoughton},
  {Strauss}, {Stubbs}, {SubbaRao}, {Szalay}, {Szapudi}, {Szokoly}, {Thakar},
  {Tremonti}, {Tucker}, {Uomoto}, {Vanden Berk}, {Vogeley}, {Waddell}, {Wang},
  {Watanabe}, {Weinberg}, {Yanny}, {Yasuda}, \& {SDSS Collaboration}}]{York00}
{York} D.~G. {et~al.}, 2000, \aj, 120, 1579

\end{thebibliography}
\bibliographystyle{mn2e}   

\appendix

\begin{figure}
  \centering
  \includegraphics[width=0.5\textwidth,angle=0.0]{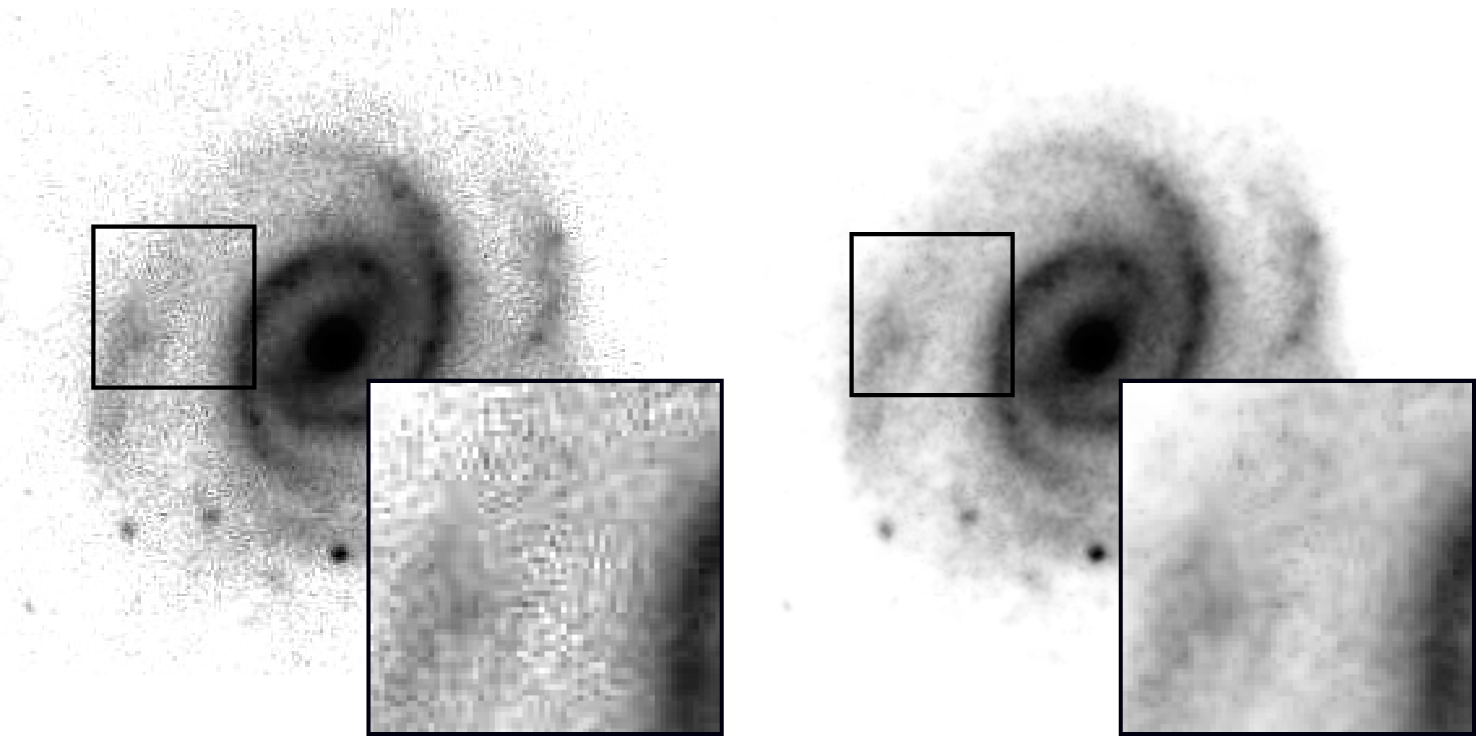}
  \caption{Image denoising and artifact removal using a stochastic sampling technique. The left panel shows the (inverted) image of a galaxy taken from the SDSS project while the right panel shows the same image after stochastic re-sampling and ensemble-averaging. The zoomed region focuses on a spiral arm with poor sampling and strong jpeg artifacts.}\label{fig:galaxy}
\end{figure}

%
\section{Image denoising and artifact reduction}\label{app:image}

In this section we describe preliminary results of a variation of the technique presented here applied to image denoising and artifact removal. We used an image downloaded as a jpeg file from the SDSS project. This image corresponds to a tenuous galaxy producing a poorly sampled image, the strong jpeg compression further introduces artifacts in regions dominated by shot noise. We generated an ensemble of images by Montecarlo-sampling the original image and computing the 2D DTFE field from the re-sampled images. The sampling density is a free parameter and was manually selected to produce visually appealing results. After $\sim 32$ realizations the image begins to converge. Our results indicate that this approach can be used as an effective adaptive filtering, not only reducing the noise in the image but also the unnatural-looking artifacts arising form the image compression while retaining even the faintest features at all scales.

\end{document}